\documentclass[aps,pra,twocolumn,showpacs,amsmath,amssymb,preprintnumbers,superscriptaddress,10pt]{revtex4-1}

\usepackage{amsmath,amssymb}
\usepackage{bm}
\usepackage[latin1]{inputenc}
\usepackage{dcolumn}
\usepackage{graphicx}
\usepackage{SIunits}

\newcommand{\qber}{\text{QBER}}
\newcommand{\f}{\text{f}}
\newcommand{\h}{\text{h}}

\begin{document}

\title{Superlinear threshold detectors in quantum cryptography}

\author{Lars Lydersen}
\email{lars.lydersen@iet.ntnu.no}
\affiliation{Department of Electronics and Telecommunications, Norwegian University of Science and Technology, NO-7491 Trondheim, Norway}
\affiliation{University Graduate Center, NO-2027 Kjeller, Norway}

\author{Nitin Jain}
\author{Christoffer Wittmann}
\affiliation{Max Planck Institute for the Science of Light, G\"{u}nther-Scharowsky-Str.\ 1/Bau 24, 91058 Erlangen, Germany}
\affiliation{Institut f\"{u}r Optik, Information und Photonik, University of Erlangen-Nuremberg, Staudtstra\ss e 7/B2, 91058 Erlangen, Germany}

\author{{\O}ystein Mar{\o}y}
\author{Johannes Skaar}
\affiliation{Department of Electronics and Telecommunications, Norwegian University of Science and Technology, NO-7491 Trondheim, Norway}
\affiliation{University Graduate Center, NO-2027 Kjeller, Norway}

\author{Christoph Marquardt}
\affiliation{Max Planck Institute for the Science of Light, G\"{u}nther-Scharowsky-Str.\ 1/Bau 24, 91058 Erlangen, Germany}
\affiliation{Institut f\"{u}r Optik, Information und Photonik, University of Erlangen-Nuremberg, Staudtstra\ss e 7/B2, 91058 Erlangen, Germany}

\author{Vadim Makarov}
\affiliation{Department of Electronics and Telecommunications, Norwegian University of Science and Technology, NO-7491 Trondheim, Norway}
\affiliation{University Graduate Center, NO-2027 Kjeller, Norway}

\author{Gerd Leuchs}
\affiliation{Max Planck Institute for the Science of Light, G\"{u}nther-Scharowsky-Str.\ 1/Bau 24, 91058 Erlangen, Germany}
\affiliation{Institut f\"{u}r Optik, Information und Photonik, University of Erlangen-Nuremberg, Staudtstra\ss e 7/B2, 91058 Erlangen, Germany}

\date{\today}

\begin{abstract}
We introduce the concept of a superlinear threshold detector, a detector that has a higher probability to detect multiple photons if it receives them simultaneously rather than at separate times. Highly superlinear threshold detectors in quantum key distribution systems allow eavesdropping the full secret key without being revealed. Here, we generalize the detector control attack, and analyze how it performs against quantum key distribution systems with moderately superlinear detectors. We quantify the superlinearity in superconducting single-photon detectors based on earlier published data, and gated avalanche photodiode detectors based on our own measurements. The analysis shows that quantum key distribution systems using detector(s) of either type can be vulnerable to eavesdropping. The avalanche photodiode detector becomes superlinear towards the end of the gate. For systems expecting substantial loss, or for systems not monitoring loss, this would allow eavesdropping using trigger pulses containing less than 120 photons per pulse. Such an attack would be virtually impossible to catch with an optical power meter at the receiver entrance.
\end{abstract}


\maketitle

\section{Introduction}
\label{sec:introduction}
Single photon detectors \cite{hadfield2009} can be regarded as essential parts of quantum information processing hardware, and are certainly crucial components in quantum key distribution (QKD) systems \cite{bennett1984,ekert1991,lo1999,shor2000,gisin2002,scarani2009}. In QKD, the communicating parties Alice and Bob exploit the properties of quantum mechanics to reveal any eavesdropping attempt by the eavesdropper Eve. The security of QKD has been proven for perfect devices \cite{lo1999,shor2000}. However, when the security of QKD is to be proven for practical systems \cite{mayers1996,inamori2007,koashi2003,gottesman2004,lo2007,zhao2008a,fung2009,lydersen2010,maroy2010}, it is necessary to construct models based on assumptions about the practical devices, and hence also about the single photon detectors.

With a few exceptions \cite{kardynal2008,lita2008}, most single photon detectors suitable for QKD systems are threshold detectors that cannot resolve the number of photons in a pulse. They rather have a binary response distinguishing between zero, and `one or more' photons, where a detection event is often referred to as a ``click''. Threshold detectors are usually characterized by their quantum efficiency $\eta$, which is the probability to detect a single photon. For multiphoton pulses, a very common assumption is that each photon within the pulse is detected individually with probability $\eta$. Then, the detection probability of a $n$-photon Fock state can be expressed as
\begin{equation}
   p_\text{det}(n) = 1 - (1 - \eta)^n.
   \label{eq:linear-detection-probability}
\end{equation}
We refer to threshold detectors with a multiphoton detection probability higher than the one given by Eq.~\eqref{eq:linear-detection-probability} as \emph{superlinear} threshold detectors. A superlinear threshold detector has a larger probability to detect multiple photons if it receives them nearly simultaneously, than if it receives each of the photons separately at different times. This effect is well known in multiphoton absorption by atoms \cite{mollow1968}, where the multiphoton absorption rate can be much higher for chaotic light than for laser light with the same mean intensity. Meanwhile for threshold detectors, superlinear response may also originate from how the entire device converts individual excitations into the macroscopic detection event.

The photon number of a coherent state follows a Poisson distribution with probability $p_n = \mu^n e^{-\mu}/n!$, where $\mu$ is the mean photon number. Therefore, if the detection probability of a $n$-photon Fock state is given by Eq.~\eqref{eq:linear-detection-probability}, a coherent state with mean photon number $\mu$ is detected with probability 
\begin{equation}
   p_\text{det} = \sum_{n = 0}^{\infty} \frac{\mu^n e^{-\mu}}{n!} p_\text{det}(n) = 1 - e^{-\mu\eta}.
   \label{eq:detection-probability-coherent-state}
\end{equation}
Note that for a coherent state with mean photon number $\mu$, a superlinear threshold detector with quantum efficiency $\eta$ will have a higher detection probability than the one given by Eq.~\eqref{eq:detection-probability-coherent-state}.

Insufficient models of single photon detectors have caused numerous security loopholes \cite{makarov2006,qi2007,lamas-linares2007,makarov2008,zhao2008,lydersen2010,lydersen2010a,lydersen2010b,wiechers2011,sauge,gerhardt2011,lydersen2011} in QKD. For instance, the time-shift attack \cite{qi2007} based on detector efficiency mismatch \cite{makarov2006} has been shown to break the security of a commercial QKD system \cite{zhao2008}. More recently, the detector control attack \cite{lydersen2010a,lydersen2010b,wiechers2011,sauge,gerhardt2011,lydersen2011} allows the eavesdropper to capture the full key without revealing her presence (via errors in the key). Specifically, the attack introduces zero quantum bit error rate (QBER). Furthermore, this attack which is based on bright illumination is implementable with current technology. Two commercial QKD systems were shown to be vulnerable to the attack \cite{lydersen2010a,lydersen2010b, wiechers2011}, and a full eavesdropper has been implemented to capture the full key of an experimental QKD system under realistic conditions \cite{gerhardt2011}. However, the power level (more than $500\,\micro\watt$) of the eavesdropper's illumination has led to discussions whether an optical power meter at the entrance of Bob can be used to detect these attacks \cite{yuan2010,lydersen2010c,yuan2011,lydersen2011d}.

In this paper we propose and analyze an attack against QKD systems with superlinear detectors (Sec.~\ref{sec:detector-control-of-non-linear-detectors}). Note that the previously published detector control attack \cite{lydersen2010a} is based on an extreme superlinear behavior of the detectors, and can therefore be considered a special case of the ``imperfect'' detector control attack presented here. Then we discuss how the attack would perform against superconducting single photon detectors \cite{gol'tsman2001,verevkin2002}, which have been reported to exhibit superlinear behavior (Section~\ref{sec:non-linearity-supderconducting-single-photon-detectors}). In Sec.~\ref{sec:non-linearity-of-the-gated-detectors} we show that APD-based gated detectors have a substantial superlinear response at the end of the gate. The superlinear behavior at the end of the gate allows eavesdropping with very faint trigger pulses \cite{lydersen2010a,lydersen2010c}. This \textit{faint after-gate attack} will be virtually impossible to catch with an optical power meter at the entrance of Bob. At least one security proof covers QKD systems with superlinear detectors \cite{maroy2010}. In Sec.~\ref{sec:countermeasures-and-proof-of-security} we show how the detector control attack relates to the security proof, and discuss possible countermeasures. Finally, we conclude in Sec.~\ref{sec:conclusion}.

\section{Theory of superlinear detector control}
\label{sec:detector-control-of-non-linear-detectors}
The core of the previously proposed detector control attacks is the following \cite{lydersen2010a}: in the Bennett-Brassard 1984 (BB84) \cite{bennett1984} family of protocols, Eve uses a random basis to measure the quantum state from Alice. Then she resends her measurement result, not as a single photon, but rather as a bright pulse, called a trigger pulse, with a carefully selected optical power. Then, if Eve uses Bob's measurement basis, her trigger pulse is \emph{always} detected by Bob. On the contrary, if Eve uses a basis not matching Bob's to measure the quantum state from Alice, her trigger pulse is \emph{never} detected. This is possible because Bob's detectors are very superlinear: for less than a factor of two ($3\,\deci\bel$) increase in trigger pulse power, the detection probability shoots from 0 to 100\%. Since Eve uses the correct basis only half of the time, the total loss between Alice and Bob is $3\,\deci\bel$. For the differential-phase-shift protocol \cite{inoue2002,inoue2003} there is no basis choice, so the same factor of two ($3\,\deci\bel$) superlinearity allows eavesdropping without extra loss \cite{lydersen2011}. The coherent one-way protocol \cite{gisin2004,stucki2005} is also vulnerable to the detector control attacks \cite{lydersen2011}, but requires a more strict relationship between the superlinearities of the detectors in the system.

The previously proposed detector control attacks allow Eve to capture the full secret key without introducing any QBER. However, Alice and Bob usually tolerate a non-zero QBER (typically less than 11\%). Therefore, Eve might introduce a small QBER without getting caught. What if the superlinearity of the detector is such that when Eve selects the right basis, the trigger pulse is detected with a \emph{high} probability, while when Eve selects the wrong basis, the trigger pulse is detected with a \emph{low} probability? One can immediately identify two consequences of this ``imperfect'' detector control attack: the non-unity detection probability when Eve uses the right basis will contribute extra to the loss. On the other hand, the non-zero detection probability when Eve uses the wrong basis will introduce a non-zero QBER.

We will here consider an active basis choice BB84 implementation using two detectors. In a passive basis choice BB84 implementation \cite{rarity1994}, Eve's trigger pulse will strike the detectors in both bases simultaneously for each bit. For this case, the QBER introduced by the attack depends on how Bob handles simultaneous clicks in both bases. Assume that Bob assigns a random bit value to these events. Then, if the probability for simultaneous clicks in both bases is non-zero, the QBER introduced by a ``imperfect'' detector control attack will be higher in a passive basis choice implementation than in an active basis choice implementation. In any case, for passive basis choice implementations, the theoretical QBER derived below can be used as a lower bound.

To calculate the QBER caused by this attack, let $p_{\f,i}$ be the detection probability in detector $i$ for the trigger pulse with full power. Likewise, let $p_{\h,i}$ be the detection probability at detector $i$ with half the power. We assume Eve resends the same power regardless of her detected bit value, that double clicks are assigned to a random bit value \cite{lutkenhaus1999}, and that Eve selects Bob's measurement basis with probability $1/2$. When Eve resends in the wrong basis and Bob has a detection, the bit value will be erroneous with probability 1/2. Therefore, the QBER caused by the ``imperfect'' detector control attack is given by
\begin{equation}
   \begin{split}
      \qber &= \frac{1}{2}\frac{\text{Bob detects and Eve used wrong basis}}{\text{Bob has a detection}} \\
         &= \frac{p_{\h,0} + p_{\h,1} - p_{\h,0}p_{\h,1}}{p_{\f,0} + p_{\f,1} + 2(p_{\h,0} + p_{\h,1} - p_{\h,0}p_{\h,1})},
   \end{split}
   \label{eq:general-QBER}
\end{equation}
where dark counts have been omitted. Errors originating from dark counts would add to the errors caused by the attack. However, in a good detector design the amount of errors from dark counts is minimized. Since we require the eavesdropper to reproduce the detection probability from normal operating conditions, the dark count probability would be minimized under attack as well. A high dark count probability, and thus a high error rate without the eavesdropper would leave the attack less room for errors to be introduced. However, an equivalent restriction on the attack is easier obtained by lowering the acceptance threshold for the QBER. Therefore, our analyses is limited to the QBER introduced by the attack, and dark counts are omitted. Assuming that both detectors have equal detection probabilities, $p_{\f,i} = p_\f$ and $p_{\h,i} = p_\h$, Eq.~\eqref{eq:general-QBER} simplifies to
\begin{equation}
      \qber = \frac{2p_\h - p_\h^2}{2p_\f + 2(2p_\h - p_\h^2)}.
   \label{eq:QBER-equal-detectors}
\end{equation}

As discussed above, the perfect detector control attack introduces $3\,\deci\bel$ loss when applied against BB84 QKD systems with active basis choice in Bob's implementation, because Eve only selects the correct basis half of the time. If $\min_i p_{\f,i} < 1$, the attack will cause an even higher loss. On the other hand, $\max_i p_{\h,i} > 0$ will reduce the loss introduced by the attack. Therefore, the transmittance $T$ when an ``imperfect'' detector control attack is applied against a BB84 QKD system with active basis choice is given by
\begin{equation}
   T = \frac{1}{4}\left(p_{\f,0} + p_{\f,1}\right) + \frac{1}{2}\left(p_{\h,0} + p_{\h,1} - p_{\h,0}p_{\h,1}\right).
   \label{eq:general-loss}
\end{equation}
Note that $T$ refers to the transmittance between Eve and Bob. If Eve uses imperfect detectors, this will add to the total loss observed by Alice and Bob. For the remainder of the paper, we simply consider Eve to use perfect detectors. Since Eve can place her detectors close to Alice, and she can use detectors with almost unity detection efficiency \cite{lita2008}, this is an acceptable assumption. If both detectors have equal probabilities, Eq.~\eqref{eq:general-loss} simplifies to 
\begin{equation}
   T = \frac{1}{2} p_\f + \frac{1}{2}\left( 2p_\h - p_\h^2\right).
   \label{eq:loss-equal-detectors}
\end{equation}
Note that in passive implementations of Bob, such as passive basis choice in BB84 \cite{rarity1994}, or in distributed phase reference protocols \cite{inoue2002,inoue2003,gisin2004,stucki2005}, there is no $3\,\deci\bel$ loss due to basis choice. Therefore, the above expression for the transmittance $T$ can be considered a lower bound also for such implementations.

In most cases, the eavesdropper can introduce substantial loss without getting noticed. With the notable exception of transition-edge sensors \cite{lita2008}, the quantum efficiency of Bob's detectors is typically about 10\% at telecom wavelengths \cite{hadfield2009}. Furthermore, an optical fiber usually exhibits a loss of about $0.2\,\deci\bel/\kilo\meter$ at $1550\,\nano\meter$ wavelength. Adding the loss owing to detector's quantum efficiency to the loss in the line at a typical distance of $50\,\kilo\meter$, Alice and Bob normally observe a total loss of $20\,\deci\bel$, corresponding to $T \sim 0.01$. In addition to this, there is loss in the optical path inside Bob's apparatus. However, Eve can always adjust the power in her trigger pulses to strike Bob's detectors with a given optical power. Therefore, by inserting her eavesdropping station into the line close to Alice's system, Eve has almost the full $20\,\deci\bel$ at her disposal. In one case, a QKD system operating with loss up to $40\,\deci\bel$ has been reported \cite{takesue2007} (but the actual, tolerable loss might be less because there is no satisfactory security proof for the protocol used in Ref.~\cite{takesue2007}). Therefore, it seems that for many QKD setups, Eve can introduce loss of more than $20\,\deci\bel$ without being revealed from the reduction in the transmittance.

\section{Superlinearity of superconducting single photon detectors}
\label{sec:non-linearity-supderconducting-single-photon-detectors}
Superconducting single photon detectors (SSPDs) based on superconducting nanowires \cite{gol'tsman2001} have been used for long-distance QKD experiments \cite{hadfield2006,takesue2007,rosenberg2009,stucki2009,liu2010}, due to their ultra low dark count rate and timing jitter. However, the need for cryogenic cooling to temperatures in the 2--4$\,\kelvin$ range has prevented them from being used in commercial QKD systems.

In SSPDs, the nanowire is cooled to the superconducting state. Then, then the nanowire is biased with a current $I_b$ slightly lower than the critical current $I_c$. Because the wire is superconducing at $I_b$, there is no voltage drop over the device. A photon incident on the nanowire can create a normally-conducting hotspot, with the effect that the whole cross-section of the nanowire becomes normally conducting. This increases the voltage over the device. Afterwards, the cooling restores superconductivity in the nanowire, and the current increases back to the bias current. This dead time is usually about $10\,\nano\second$. The biasing current $I_b$ can be adjusted for a trade-off between high detection efficiency and low dark count rate.

\begin{figure}[t!]
   \includegraphics[width=8.6cm]{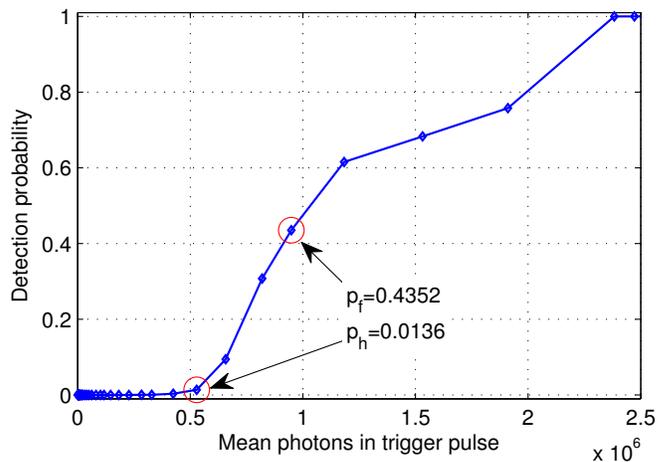}
   \caption{(Color online) The detection probability versus mean photon number in the trigger pulse for the SSPD in Ref.~\cite{verevkin2002}, at $1550\,\nano\meter$ and $I_b = 0.8I_c$. Count rates were extracted from Fig.~1 in Ref.~\cite{verevkin2002}, and divided by the pulse repetition frequency of $82\,\mega\hertz$ to obtain the detection probability. The red circled data points were used to calculate the QBER and the transmittance from an attack.}
   \label{fig:detection-probability-sspd} 
\end{figure}

Already in the first systematic investigation of the detection efficiency of SSPDs \cite{verevkin2002}, superlinear behavior due to multiphoton absorption mechanisms was reported. The superlinear behavior is wavelength dependent, and is substantial at $1550\,\nano\meter$, which is the wavelength suitable for long-distance experiments. Figure~\ref{fig:detection-probability-sspd} shows the detection count data for $1550\,\nano\meter$ extracted from Fig.~1 in Ref.~\cite{verevkin2002}, processed as detection probability (count rate/trigger pulse rate), and plotted on a linear scale. The SSPD was biased at $I_b/I_c = 0.8$. The superlinear behavior is suitable for eavesdropping in QKD: by increasing the photon number, the detection probability increases sharply. Using trigger pulses containing $10^6$ photons per pulse, Eq.~\eqref{eq:QBER-equal-detectors} predicts a QBER of less than 3\%, and Eq.~\eqref{eq:loss-equal-detectors} predicts a transmittance $T > 0.20$ (assuming reasonable errors in extracting the numerical data from the plot in Ref.~\cite{verevkin2002}). Therefore, a QKD system using this SSPD would clearly be vulnerable to a detector control attack.

Judging by the low detection probability at one photon per pulse for this SSPD, the QKD experiments would use a higher bias current to get better sensitivity. Unfortunately, few publications seems to report the detection probabilities for pulses above the single photon level, especially for $1550\,\nano\meter$ wavelength. The available literature shows that SSPDs are less superlinear at shorter wavelengths \cite{verevkin2002}, and also less superlinear at higher bias currents \cite{akhlaghi2009}. However, note that any superlinear detector response must be handled in the security proof. Therefore, the reported data on SSPDs \cite{verevkin2002,akhlaghi2009} clearly shows that such a security proof is necessary for QKD systems using SSPDs.

\section{Superlinearity of gated APD-based detectors}
\label{sec:non-linearity-of-the-gated-detectors}
The gated APD-based detectors in the QKD system Clavis2 by ID~Quantique exhibit substantial superlinear behavior far after the gate \cite{wiechers2011}, or when blinded by bright illumination \cite{lydersen2010a,lydersen2010b}. However, as pointed out before \cite{lydersen2010a,yuan2010}, the bright trigger pulses might be revealed by an optical power meter at the entrance of Bob. Here, we show that at the end of the gate, when the APD is biased close to the breakdown voltage, the superlinear response allows Eve to use very faint trigger pulses.

The detection probability during the gate was measured as follows: The gated InGaAs detectors in the QKD system Clavis2 were run with factory settings, but with the gating frequency reduced from $5\,\mega\hertz$ to $98\,\kilo\hertz$. The reduced frequency corresponds to the factory frequency with a detection in every gate, and afterpulse blocking (forced $10\,\micro\second$ deadtime after detection events to reduce dark counts) enabled. A short-pulsed laser (see Appendix~\ref{sec:pulse-shape-of-id300} for the pulse shape) was attenuated to the appropriate mean photon number, and connected directly to the fiber pigtail of each detector. Then, the laser pulse was scanned through the gate in steps of $25\,\pico\second$, and the detection probability was recorded in each step. The ``quantum efficiency'' $\eta$ was measured by applying a coherent state $\mu = 1$, and solving $\eta$ from Eq.~\eqref{eq:detection-probability-coherent-state}. In fact, the detector is slightly superlinear, but a coherent state with $\mu=1$ \footnote{The coherent state with $\mu=1$ was obtained by shining much brighter pulses into a power meter and calculating the energy per pulse. Then a controlled amount of optical attenuation was introduced to decrease the energy level of each pulse to $\mu=1$.} contains only a small fraction of multiphoton pulses.

\begin{figure}[t!]
   \includegraphics[width=8.6cm]{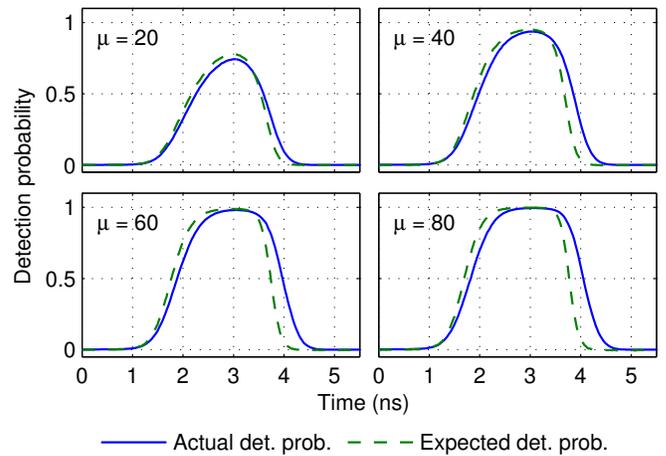}
   \caption{(Color online) The measured detection probability for a coherent state with $\mu=20$, 40, 60 and 80, compared to the expected detection probabilities predicted by Eq.~\eqref{eq:detection-probability-coherent-state}. Data points are $25\,\pico\second$ apart. Data for detector~0 is shown; detector~1 behaved very similarly. When the mean photon number $\mu$ is increased, deviation between the expected detection probabilities and the actual measured detection probabilities increases, especially at the end of the gate. See also Fig.~\ref{fig:non-linearity-vs-mean-photon-number}.}
   \label{fig:detection-probability-coherent-state} 
\end{figure}

Once the quantum efficiency $\eta$ is known, Eq.~\eqref{eq:detection-probability-coherent-state} can be used to calculate the expected detection probability for a coherent state with any mean photon number, assuming that each photon is detected individually. Figure~\ref{fig:detection-probability-coherent-state} shows the detection probability of a coherent state for various mean photon numbers predicted by Eq.~\eqref{eq:detection-probability-coherent-state}, compared to the actual detection probabilities measured in our experiment.

The measurement data matches the expected detector response fairly well until the falling edge of the gate. There, the measured detection probability becomes superlinear. One possible explanation for this could be the following: an avalanche, started by a photon in a localized spot, laterally spreads over time to encompass the entire junction area of the APD \cite{spinelli1997}. For detection events before the falling edge of the gate, the avalanche has sufficient time to spread and therefore the current reaches the same amplitude regardless of the number of photons absorbed in the APD \cite{kardynal2008}. At the end of the gate, an avalanche from a single photon absorption does not have sufficient time to spread to the entire junction area, and therefore only causes a small current insufficient of crossing the comparator threshold. However, multiple photon absorptions in different spots across the junction can start multiple small avalanches that together provide enough current to cross the comparator threshold. This is exactly the process exploited to make photon number resolving APD-based detectors \cite{kardynal2008}. Avalanche spreading assisted by secondary photons re-emitted by the APD, has already been used to explain avalanche development \cite{spinelli1997,cova2004}. Similarly, multiple photon absorptions caused by the multiphoton pulse could speed up the avalanche development.

For the gated APD-based detectors, the superlinear response can be exploited in a faint version of the after-gate attack \cite{wiechers2011}. From Eve's perspective, the original after-gate attack has some drawbacks. The attack may cause a substantial amount of errors in the key, because the bright pulses cause afterpulses with a random bit value. Furthermore, in principle, an optical power meter can be used to catch Eve's bright pulses. Also, removing gates randomly or as a part of afterpulse blocking (to avoid excessive dark counts) would reveal the attack because the trigger pulses would cause clicks regardless of the presence of a gate. Then, detection events without a gate applied would indicate the presence of the eavesdropper. Similarly, it has been noted that in the original after-gate attack could be countered by ignoring detection events outside the gate \cite{yuan2011}, while for this faint after-gate attack, the detections happen within the gate \cite{lydersen2011d}.

\begin{figure}[t!]
   \includegraphics[width=8.6cm]{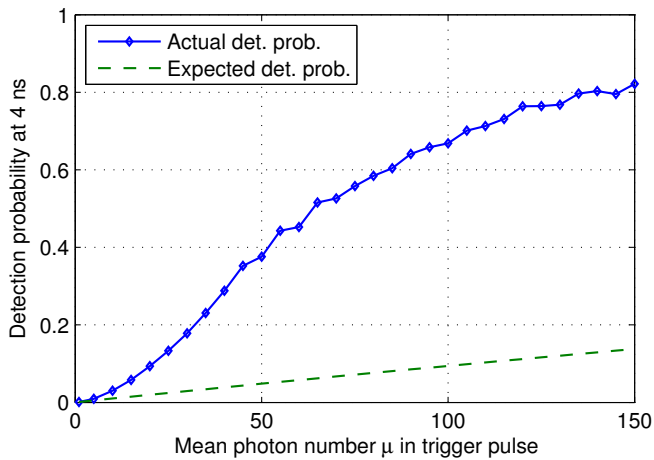}
   \caption{(Color online) Detection probability at the falling edge of the gate (at the $4\,\nano\second$ point in Fig.~\ref{fig:detection-probability-coherent-state}). For $\mu < 40$, the shape of the actual detection probability is clearly superlinear, in contrast to the nearly linear expected detection probability.}
   \label{fig:non-linearity-vs-mean-photon-number}
\end{figure}

As discussed in Sec.~\ref{sec:detector-control-of-non-linear-detectors}, having a ``high'' detection probability for a given trigger pulse power, and a ``low'' detection probability for a $3\,\deci\bel$ dimmer trigger pulse is suitable for Eve's attack. Figure~\ref{fig:non-linearity-vs-mean-photon-number} shows the measured and expected detection probability at a single point at the falling edge of the gate. For less than 40 photons per trigger pulse, the APDs clearly exhibit superlinear response in favor of the eavesdropper.

\begin{figure}[t!]
   \includegraphics[width=8.6cm]{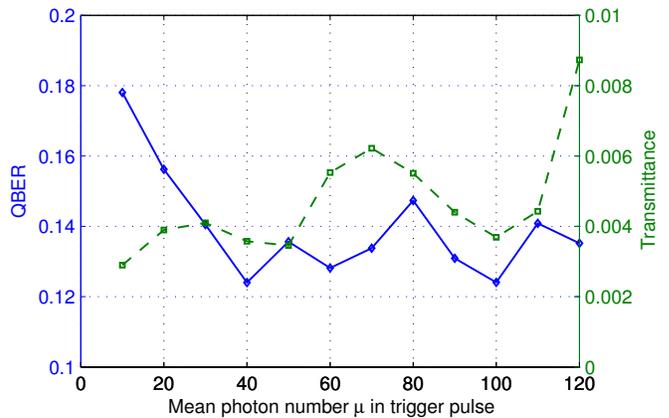}
   \caption{(Color online) The minimum QBER (solid curve) caused by trigger pulses with various mean photon number calculated from Eq.~\eqref{eq:QBER-equal-detectors} and the corresponding transmittance (dashed curve) calculated from Eq.~\eqref{eq:loss-equal-detectors}. The data contains some noise due to fluctuations in applied power and/or fluctuations in the detection efficiency.}
   \label{fig:mean-photon-number-vs-qber}
\end{figure}

The detection probability curve of the detector~0 (the results are very similar for detector~1) was used when calculating QBER and transmittance from Eqs.~\eqref{eq:QBER-equal-detectors} and \eqref{eq:loss-equal-detectors}. Figure~\ref{fig:mean-photon-number-vs-qber} shows the resulting  QBER and the corresponding transmittance for various mean photon numbers in the trigger pulse, when the trigger pulse timing was optimized to minimize the QBER. The data indicates that a faint after-gate attack could cause a QBER around 13\% with a transmittance of about 0.005, corresponding to $23\,\deci\bel$ loss (for instance, for $\mu=40$, $p_f = 0.0054$ and $p_h = 0.00089$ at the point $4.525\,\nano\second$ in Fig.~\ref{fig:detection-probability-coherent-state}). As discussed in Sec.~\ref{sec:detector-control-of-non-linear-detectors}, this transmittance corresponds to Bob's detectors having 10\% quantum efficiency, a line loss corresponding to about $50\,\kilo\meter$ of fiber and another $3\,\deci\bel$ loss in Bob's apparatus, which are reasonable values.

While most QKD systems do not accept QBER above 11\% \cite{shor2000}, there are post-processing protocols which accept QBER up to 20\% \cite{chau2002}. Also note that the QBER introduced by the attack may be significantly lower with yet shorter trigger pulses, since they would better resolve the superlinear behavior at the falling edge of the gate. A relatively wide pulse we're using (Appendix~\ref{sec:pulse-shape-of-id300}) arrives at both linear and superlinear regions of the gate. Therefore the superlinear response to it must be less than that to a narrower pulse arriving only at the most superlinear point in the gate.

The detectors in Clavis2 have been shown to exhibit detection efficiency mismatch \cite{makarov2006,zhao2008,jain}. Therefore, in the general case one would have to use different timings and/or different powers depending on the bit value, to avoid skewing the bit value distribution in the raw key. Also, the superlinearity could be exploited in other attacks, such as the faked-state attack \cite{jain,makarov2005} and conventional quantum attacks, to make them more efficient \cite{lydersen2010}.

ID~Quantique has been notified about this loophole prior to the submission of the manuscript.

\section{Countermeasures and proof of security}
\label{sec:countermeasures-and-proof-of-security}
The security of QKD systems with arbitrary non-linearities in Bob's system, and therefore superlinear threshold detectors has already been proved \cite{maroy2010}. Without source imperfections and with symmetry in the two bases, the secret key rate is given by \cite{maroy2010}
\begin{equation}
   R \geq -h(\qber) + \eta(1-h(\qber)),
   \label{eq:maroy-rate}
\end{equation}
where $h(\cdot)$ is the binary entropy function, and $\eta$ is the smallest detection probability of a non-vacuum state. If one further assumes that the probability to detect a multiphoton state is higher than a single photon, $\eta$ is simply the quantum efficiency (the probability to detect a single photon). 

\begin{figure}[t!]
   \includegraphics[width=8.6cm]{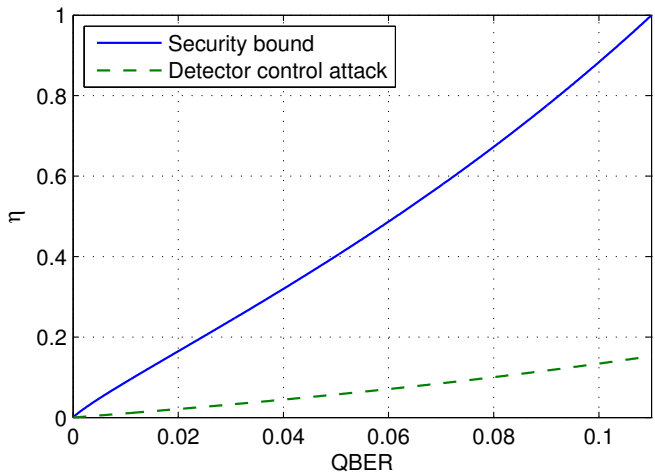}
   \caption{(Color online) Comparison of the detector control attack and the bound from the security proof \cite{maroy2010}. The region to the left of the security bound curve (solid curve) allows extraction a secure key. The region to the right of the detector control attack curve (dashed curve) is clearly insecure, because the attack presented in Sec.~\ref{sec:detector-control-of-non-linear-detectors} can be applied. The region between the curves should be assumed insecure.}
   \label{fig:comparison-maroy2010}
\end{figure}

As for the detector control attack, let us assume the worst-case superlinearity, namely that a single photon is detected with probability $\eta$, while a two-photon state is detected with probability 1. Then, Eve can use trigger pulses with two photons, and Eq.~\eqref{eq:QBER-equal-detectors} simplifies to
\begin{equation}
   \qber = \frac{2\eta - \eta^2}{2 + 2(2\eta - \eta^2)}.
   \label{eq:qber-maximum-non-linearity}
\end{equation}
Figure~\ref{fig:comparison-maroy2010} shows Eq.~\eqref{eq:maroy-rate} for $R=0$ and Eq.~\eqref{eq:qber-maximum-non-linearity}, comparing the ``imperfect'' detector control attack with the bounds derived in the security proof \cite{maroy2010}. It shows that a sufficiently high detection probability, and thus quantum efficiency allows extraction of secret key regardless of any superlinear detector response. For instance, if the QBER is 5\%, a quantum efficiency $\eta > 0.4$ allows the extraction of secret key. Note that a high quantum efficiency does not remove the superlinear effect, but then the security proof makes it possible to remove any knowledge Eve could have obtained exploiting the superlinear response, by (a large amount of) extra privacy amplification \cite{bennett1995}.

For gated systems, one possible countermeasure might be bit-mapped gating \cite{lydersen2011a}. Then, the basis selector is used to randomize all detection events outside the center of the gate. Therefore, trigger pulses timed at the falling edge of the gate would cause random detections and thus a QBER of 50\%. This would reveal Eve's presence. However, the security analysis for bit-mapped gating requires that each photon is detected individually during the transition of the basis selector. In practice, this means that the detectors must have a detection probability given by Eq.~\eqref{eq:detection-probability-coherent-state} in the center of the gate. Figure~\ref{fig:detection-probability-coherent-state} shows that this is nearly the case. It might be possible to detect each photon completely individually in the middle of the gate by expanding the gate, or by shaping the applied electrical gate appropriately.

\section{Summary and conclusion}
\label{sec:conclusion}
In this paper we have analyzed the security of QKD systems using superlinear threshold detectors. The detector control attack previously reported \cite{lydersen2010a} is based on very superlinear detection probability: when the amplitude of the trigger pulses is increased, the detection probability sharply increases from 0 to 100\%. This allows eavesdropping the full key without causing any errors, the only side effect is $3\,\deci\bel$ total loss. Here, the detector control attack is generalized to moderately superlinear detectors by accepting a limited amount of errors in the key, and/or accepting a higher loss. Note that in practice, a total loss of about $20\,\deci\bel$ may be tolerable, as discussed in Sec.~\ref{sec:detector-control-of-non-linear-detectors}.

Nanowire SSPDs \cite{gol'tsman2001} have been reported to have superlinear detection probability \cite{verevkin2002}. We have shown that by carefully selecting the trigger pulse amplitude, an eavesdropper would introduce a QBER of less than 3\% when attacking the SSPD in Ref.~\cite{verevkin2002}. The total loss caused by the eavesdropping would be less than $6\,\deci\bel$. Therefore, a QKD system using this detector would clearly be insecure.

Figures~\ref{fig:detection-probability-coherent-state} and \ref{fig:non-linearity-vs-mean-photon-number} show that the response of the APD-based gated detector is superlinear at the falling edge of the gate. Therefore, it is possible to attack the gated detectors with faint trigger pulses, with less than 120 photons per pulse. From the measurements, the attack would cause a QBER of about 13\% and about $23\,\deci\bel$ loss. Most QKD systems do not accept a QBER above 11\% \cite{shor2000}, but there are post-processing protocols allowing a QBER up to 20\% \cite{chau2002}. Furthermore, we suspect that both the QBER and the loss could be reduced by using shorter trigger pulses \footnote{Which we unfortunately did not have at our disposal for this experiment.}. Finally, even if the attack is not directly applicable to some QKD systems due to the QBER and/or loss threshold, the superlinear response of the APD-based detector shows that ordinary security proofs  no longer apply to these systems. Therefore, these systems must use advanced security proofs to bound and remove Eve's partial knowledge from the moderate superlinear response.

The faint after-gate attack does not suffer from the limitations of the original after-gate attack \cite{wiechers2011}. In the faint after-gate attack, the afterpulsing is negligible. Furthermore, with less than 120 photons per pulse, the trigger pulses should be nearly impossible to catch with an optical power meter at the entrance of Bob. Also, removing gates randomly or due to after-pulse blocking will not expose the attack \cite{wiechers2011} since such trigger pulse will not cause a click unless there is a gate present. Furthermore, the timing of the trigger pulse detection will be very similar to a normal detection inside the gate, and therefore difficult to discard based on timing \cite{yuan2011}.

If the detectors have an increasing detection probability for increasing photon number, a sufficiently high quantum efficiency makes it possible to remove Eve's knowledge using privacy amplification \cite{bennett1995}. For gated APD-based detectors, bit-mapped gating \cite{lydersen2011a} can be used if each photon is detected individually in the center of the gate.

Quantum key distribution has been proven secure for all future, so currently the challenge is to make a secure implementation. We believe that weeding out loopholes caused by the implementation is a necessary step towards achieving practical secure QKD, and that this work is crucial because it fully exposes the nature of the detector control attack.

\begin{acknowledgments}
We acknowledge ID~Quantique's valuable assistance in this project. We also acknowledge useful discussions with M.\ Akhlaghi. This work was supported by the Research Council of Norway (grant no.\ 180439/V30), DAADppp mobility program financed by NFR (project no.\ 199854), and DAAD (project no.\ 50727598). L.L.\ and V.M.\ acknowledge travel support from the Institute for Quantum Computing, Waterloo, Canada.
\end{acknowledgments}

\begin{figure}[t!]
   \includegraphics[width=8.6cm]{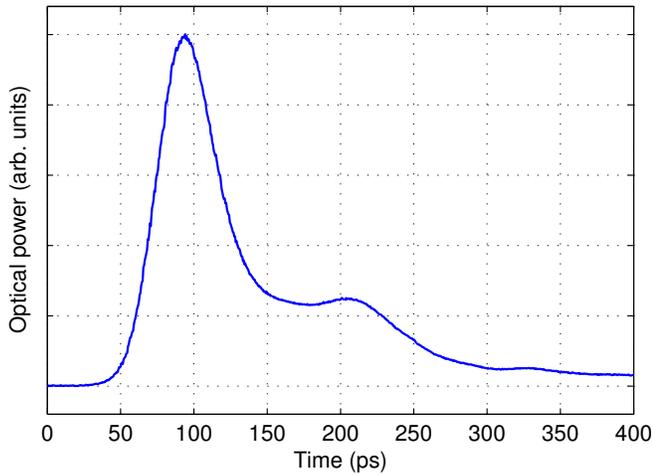}
   \caption{(Color online) Pulse shape of the id300 short-pulsed laser, measured with a $45\,\giga\hertz$ optical probe on a $12.5\,\giga\hertz$ sampling oscilloscope at a pulse repetition rate of $100\,\kilo\hertz$.}
   \label{fig:pulse-shape-of-id300} 
\end{figure}

\appendix
\section{Pulse shape of id300}
\label{sec:pulse-shape-of-id300}

Figure~\ref{fig:pulse-shape-of-id300} shows the pulse shape of the id300 short-pulsed laser \cite{idq-id300}. This is the particular laser sample used in this experiment; other samples of this laser model may have a different pulse shape.

\bibliography{bibtex_library}

\end{document}